\documentstyle[12pt,psfig]{article}

\def\al{\alpha}

\def\ga{\gamma}
\def\de{\delta}
\def\ep{\epsilon}

\def\et{\eta}
\def\th{\theta}

\def\ka{\kappa}
\def\la{\lambda}

\def\si{\sigma}

\def\ph{\phi}

\def\ch{\chi}
\def\ps{\psi}

\def\Ga{\Gamma}

\def\Om{\Omega}

\def\cC{{\cal C}}

\def\cl{{\cal L}}
\def\cT{{\cal T}}
\def\fr#1#2{{{#1} \over {#2}}}
\def\prt{\partial}
\def\pt#1{\phantom{#1}}

\def\half{{\textstyle{1\over 2}}}
\def\frac#1#2{{\textstyle{{#1}\over {#2}}}}
\def\lsim{\mathrel{\rlap{\lower4pt\hbox{\hskip1pt$\sim$}}
    \raise1pt\hbox{$<$}}}
\def\gsim{\mathrel{\rlap{\lower4pt\hbox{\hskip1pt$\sim$}}
    \raise1pt\hbox{$>$}}}
\def\sqr#1#2{{\vcenter{\vbox{\hrule height.#2pt
         \hbox{\vrule width.#2pt height#1pt \kern#1pt
         \vrule width.#2pt}
         \hrule height.#2pt}}}}

\def\lrprtnu{\stackrel{\leftrightarrow}{\partial^\nu}}

\def\lrDnu{\stackrel{\leftrightarrow}{D_\nu}}
\def\lrDnuupper{\stackrel{\leftrightarrow}{D^\nu}}
\def\slash#1{\not\hbox{\hskip -2pt}{#1}}

\def\mn{{\mu\nu}}

\newcommand{\beq}{\begin{equation}}
\newcommand{\eeq}{\end{equation}}
\newcommand{\bea}{\begin{eqnarray}}
\newcommand{\eea}{\end{eqnarray}}
\newcommand{\rf}[1]{(\ref{#1})}
 
\renewenvironment{thebibliography}[1]
 { \rm
   \begin{list}{\arabic{enumi}.}
    {\usecounter{enumi} \setlength{\parsep}{0pt}
     \setlength{\itemsep}{3pt} \settowidth{\labelwidth}{#1.}
     \sloppy
    }}{\end{list}}
 
\begin{document}
\titlepage
 
\baselineskip=20pt

\begin{flushright}
{IUHET 432\\}
{April 2001\\}
\end{flushright}

\vglue 1cm

\begin{center}
{{\bf CROSS SECTIONS AND LORENTZ VIOLATION
}
\vglue 1.0cm
{Don Colladay$^a$ and V.\ Alan Kosteleck\'y$^b$
\\} 

\bigskip
{\it $^a$New College, University of South Florida,\\}
{\it Sarasota, FL 34243, U.S.A.\\}

\bigskip
{\it $^b$Physics Department, Indiana University\\}
{\it Bloomington, IN 47405, U.S.A.\\}
}
\vglue 0.8cm
  
\end{center}

\bigskip

{\rightskip=2pc\leftskip=2pc\noindent
The derivation of cross sections and decay rates 
in the Lorentz-violating standard-model extension 
is discussed.
General features of the physics are described,
and some conceptual and calculational issues are addressed.
As an illustrative example,
the cross section for the
specific process of electron-positron pair annihilation 
into two photons is obtained.
}

\vskip 1 cm

\baselineskip=20pt
\newpage

\noindent 
{\it 1.\ Introduction.}
The possibility of small violations 
of Lorentz invariance in quantum field theory
is theoretically and experimentally viable
\cite{cpt98}.
At the level of the standard model,
a general Lorentz-violating standard-model extension
is known
\cite{ck}.
Its lagrangian consists of all possible terms
involving standard-model fields
that are observer Lorentz scalars,
including terms having coupling coefficients with Lorentz indices.
Gauge invariance is usually imposed.
At low energies,
the relevant operators are renormalizable
and are all given in Ref.\ \cite{ck}.
The domain of validity of the renormalizable terms
in the fermion sector
is known to be below the Planck scale or,
for some operators,
below the geometric mean of the low-energy scale and the Planck scale
\cite{kle}.
Above this scale,
the nonrenormalizable terms would play a crucial role
in maintaining causality and stability of the theory.

Various experiments have placed bounds on
parameters in the standard-model extension,
including
comparative tests of quantum electrodynamics (QED)
in Penning traps \cite{bkr,gg,hd,rm},
spectroscopy of hydrogen and antihydrogen \cite{bkr2,dp},
measurements of muon properties
\cite{bkl,vh},
clock-comparison experiments
\cite{ccexpt,kla,lh,db},
observations of the behavior of a spin-polarized torsion pendulum
\cite{bk,bh},
measurements of cosmological birefringence \cite{cfj,ck,jk,pvc},
studies of neutral-meson oscillations
\cite{kpcvk,k99,bexpt},
and observations of the baryon asymmetry \cite{bckp}.
However,
relatively little is known about the implications
of the theory for scattering experiments or particle decays.
For instance,
although a consistent quantization and the
associated Feynman rules are known,
no complete calculation of a scattering process
in the context of the standard-model extension
has been performed to date.

In the present work,
we describe a general procedure for the calculation of cross sections
and decay rates in the standard-model extension.
Some of the usual concepts and tools 
are based on Lorentz invariance,
so alternative procedures are needed. 
As an illustrative example,
we obtain the cross section for the
specific process of relativistic electron-positron 
pair annihilation into two photons,
using the general Lorentz-violating extension 
of QED.
The presence of Lorentz violation introduces new physical features,
as shown below.
For example,
the violation of rotational invariance 
implies that the scattering cross section depends 
on the orientation of the colliding beams,
which in turn produces in physical observables
a sidereal-time dependence arising from the Earth's rotation.

\vglue 0.4cm
\noindent 
{\it 2.\ General considerations.}
We begin with some general issues 
affecting the calculation of cross sections and decay rates 
in the standard-model extension.
For simplicity in what follows,
specific examples are restricted to the theory
of electrons, positrons, and photons
in the presence of Lorentz and CPT violation.
This QED extension 
is a limit of the full standard-model extension,
and the methods used below apply directly 
to the complete theory.

It is useful first to recap the definition and some properties 
of the general Lorentz- and CPT-violating QED extension.
Its lagrangian is
\cite{ck}
\beq
\cl = 
\half i \overline{\ps} \Ga^\nu \lrDnu \ps 
- \overline{\ps} M \ps
- \frac 1 4 F_{\mu\nu}F^{\mu\nu} 
+ \half {(k_{AF})}^\ka \ep_{\ka\la\mu\nu} A^\la F^{\mu\nu} 
-\frac 1 4 {(k_F)}_{\ka\la\mu\nu} F^{\ka\la}F^{\mu\nu} .
\label{lag}
\eeq
In this equation,
\beq
{\Ga}^{\nu}={\ga}^{\nu}+c^{\mu \nu}
{\ga}_{\mu}+d^{\mu \nu}{\ga}_{5} {\ga}_{\mu}
+e^{\nu}+if^{\nu}{\ga}_{5}
+\frac{1}{2}g^{\la \mu \nu}
{\sigma}_{\la \mu},
\label{Gam}
\eeq
and
\beq
M=m+a_{\mu}{\ga}^{\mu}+b_{\mu}{\ga}_{5}
{\ga}^{\mu}+\frac{1}{2}H^{\mu \nu}
{\sigma}_{\mu \nu}.
\label{M}
\eeq 
Here,
$m$ is the electron mass,
while the quantities
$a_\mu$, $b_\mu$, $c_{\mu\nu}$, $d_{\mu\nu}$,
$e_\mu$, $f_\mu$, $g_{\la\mu\nu}$, $H_{\mu\nu}$,
${(k_{AF})}_\ka$, ${(k_F)}_{\ka\la\mu\nu}$
are real coefficients controlling the Lorentz violation.

The Lorentz-violating terms in the lagrangian \rf{lag}
produce effects both in relativistic quantum mechanics 
and in quantum field theory
\cite{ck,kle}.
In quantum theory,
the presence of extra time derivatives
in the quadratic fermion component of Eq.\ \rf{lag}
implies that the time evolution of $\ps$ is unconventional,
so the asymptotic states associated with $\ps$ 
cannot be directly identified with physical free-particle states.
To avoid these interpretational difficulties, 
the extra time-derivative factors
can be eliminated via a spinor redefinition 
\cite{bkr}
\beq
\ps = A \ch,
\label{redef}
\eeq
where $A$ is an invertible matrix satisfying
$A^\dagger \ga^0 \Ga^0 A = 1$.
This redefinition leaves the physics unchanged 
but yields the conventional Schr\"odinger time evolution for $\ch$,
$i\prt_0 \ch = H \ch$.
The quanta created by $\ch$
can therefore be identified with physical particles.
The modified hamiltonian $H$ is hermitian and is given by
\beq
H = \ga^{0} \overline A ( - i \Ga_{j} D^{j} +  M) A ,
\eeq
where $\overline A \equiv \ga^{0} A^{\dagger} \ga^{0}$.
Details about the existence of the spinor redefinition 
and the properties of the matrix $A$
can be found in Ref.\ \cite{kle}.

Since $A$ is chosen by definition to eliminate time derivatives,
its form depends on the specification of the time coordinate
and hence on the choice of observer inertial frame.
This induces a noncovariant relationship 
between $\ch$ in the chosen frame
and the corresponding physical spinor in another frame
boosted relative to the first.
Awareness of this feature of the spinor redefinition 
is crucial in calculating scattering cross sections
or decay rates,
since several of the standard procedures
rely upon conversions between various special frames
such as the rest frame, 
the center-of-momentum frame,
or the laboratory frame.
Transformations between inertial frames 
can still be implemented,
but the additional complexity 
arising from the associated transformations of $A$ 
makes it easier in practice to adopt calculational methods
that avoid frame changes.
We emphasize that all these complications are purely technical,
since the physics is guaranteed to be independent
of the choice of observer inertial frame
by virtue of observer Lorentz invariance
\cite{ck}.

The calculation of cross sections or decay rates 
parallels the conventional approach with some modifications.
The $S$-matrix element and 
the corresponding transition probability per unit time and volume 
can be calculated from Feynman diagrams in the standard way,
but with appropriately modified Feynman rules
\cite{ck}.
Associated to the lagrangian \rf{lag}
is a conserved canonical energy-momentum tensor,
and the corresponding energy-momentum 4-vector $p^\mu$
is conserved at the vertices of the Feynman diagrams as usual.
This generates the standard overall momentum $\de$-function factor.
For external legs on a diagram,
spinor solutions of the modified Dirac equation must be 
derived and used.
Internal lines in Feynman graphs 
are associated with a modified propagator,
which in the case of fermions in the QED extension
takes the form
\beq
S_{F}(p) = 
\fr 1 
{\ga^{0} E - \overline A \vec \Ga A \cdot \vec p -
\overline A M A} .
\label{genprop}
\eeq
This propagator for $\ch$ is appropriate for 
calculations of physical observables in scattering processes.
Note that it differs from the propagator for $\ps$,
which has been used to study microcausality of the theory
\cite{kle}.

As usual,
to obtain the cross section or decay rate,
the transition probability per unit time and volume
must be divided by a factor $F$
accounting for properties of the initial state.
For example,
for the scattering between two beams of particles,
this factor is normally defined as 
the product of the beam densities $N_1$, $N_2$
and the modulus of the beam velocity difference
$|\vec v_1-\vec v_2|$:
\beq
F = N_1 N_2 |\vec v_1-\vec v_2|.
\label{F}
\eeq
In the conventional case,
Lorentz invariance and the momentum-velocity relation
$\vec p = E \vec v$ are then often exploited
to write $F$ in the covariant form 
$4[(p_1 \cdot p_2)^2 - m^4]^{1/2}$,
with the field normalizations chosen 
so that the densities are $N_1 = 2E_1$, $N_2 = 2 E_2$.

In the present Lorentz-violating case,
Eq.\ \rf{F} still holds by definition
but some care is required in its application
because the momentum-velocity relation is modified
and because the frame-dependence of 
the field redefinition \rf{redef}
complicates the transformation of $F$ between frames.
It is therefore most convenient 
to calculate the entire cross section 
in a single inertial frame.
In the chosen frame,
the factor $F$ is obtained from Eq.\ \rf{F}.
The beam densities are derived using the time component 
of the conserved current,
as usual.   
The beam velocities are calculated from the dispersion relation,
via the definition of the group velocity 
of a wave packet with momentum $\vec p$:
\beq
\vec v_g \equiv \vec \nabla_{p} E(\vec p) .
\label{vg}
\eeq
The reader is reminded that the group velocity $\vec v_g$
and the phase velocity 
$\vec v_p \equiv E \hat p /|\vec p|$ 
typically differ in orientation in the Lorentz-violating case
\cite{ck}.
Finally,
the result for $F$ can be combined with the transition probability 
per unit time and volume to yield the physical cross section 
for any given process in the presence of Lorentz violation.

\vglue 0.4cm
\noindent 
{\it 3.\ Relativistic electrons and positrons.}
A standard class of experiments in QED
involves relativistic electron-positron collisions.
In this case,
the treatment of the fermion sector 
of the lagrangian \rf{lag} simplifies,
as is discussed next.

In the relativistic limit, 
effects from the nonderivative couplings in Eq.\ \rf{lag}
associated with the coefficients
$a_\mu$, $b_\mu$, $H_{\mu\nu}$
are suppressed relative to other Lorentz-violating terms.
This suppression occurs because 
the effects of the derivative couplings on the dispersion relation
grow with momentum,
while those of the nonderivative couplings are
momentum independent.
We emphasize that even at high energies
the effects of the derivative couplings remain small 
relative to the conventional behavior 
from the fermion kinetic term,
since this also grows with momentum.
The point is rather that effects from 
Lorentz-violating derivative couplings 
dominate those from nonderivative ones,
so at relativistic momenta it is reasonable to neglect
the effects of the coefficients
$a_\mu$, $b_\mu$, $H_{\mu\nu}$.
Note,
however,
that caution is necessary in considering
effects above the ultrarelativistic scale 
determined by the geometric mean $\sqrt{mM_P}$
of $m$ and the Planck mass $M_P$,
since above this scale nonrenormalizable corrections
to the lagrangian \rf{lag}
become essential to maintain causality and stability 
in the quantum theory
\cite{kle}.
Other effects may also arise at the string scale
\cite{kps}.
In the present work,
we limit attention to the relativistic regime 
lying well above $m$ but below $\sqrt{mM_P}$.

To simplify further the analysis here,
we disregard possible effects from the CPT-violating coefficients
$e_\mu$, $f_\mu$, $g_{\la\mu\nu}$.
These are zero in the exact QED limit of the standard-model extension
because they are incompatible
with the full SU(2)$\times$U(1) gauge symmetry,
but in practice they might be induced 
through various radiative corrections
\cite{ck}.
Also,
in most applications,
the colliding electrons and positrons are unpolarized.
For unpolarized beams,
effects from $d_{\mu\nu}$ average to zero
because the corresponding field term
contains a $\ga_5$ coupling,
which induces a correction of opposite sign 
for $\ch_{L}$ and $\ch_{R}$.

The only remaining coefficient is $c_{\mu\nu}$,
and the corresponding field operator produces 
the dominant Lorentz-violating effects 
in relativistic collisions of unpolarized electrons and positrons.
The effective lagrangian for the fermions
is therefore a subset of the QED extension \rf{lag}:
\beq
\cl = \half i (\et_{\mu\nu} + c_{\mu\nu}) \overline{\ps} \ga^{\mu} 
(\lrprtnu + 2 i q A^{\nu}) \ps
- m \overline{\ps} \ps .
\label{pslag}
\eeq
Without loss of generality, 
$c_{\mu\nu}$ can be taken traceless.
To leading order in $c_{\mu\nu}$,
the field redefinition \rf{redef} is explicitly found to be 
\beq
\psi \equiv A \chi = (1 - \half c_{\mu 0} \gamma^{0} \gamma^{\mu})\chi .
\label{redefc}
\eeq
The lagrangian in terms of $\chi$ becomes 
\beq
\cl = 
\half i \tilde{\et}_{\mn} \overline \chi \gamma^{\mu}\lrDnuupper \chi 
- \tilde{m} \overline \chi \chi ,
\label{lagc}
\eeq
where we have introduced the convenient notation
\bea
\tilde m &\equiv& m(1 - c_{00}),
\nonumber\\
\tilde{\et}_{\mn} &\equiv& \et_{\mn} + \cC_{\mn}, 
\nonumber\\
\cC_{\mn} &\equiv &
c_{\mu\nu} - c_{\mu 0} \et_{0 \nu} 
+ c_{\nu 0} \et_{0 \mu} - c_{00} \et_{\mu\nu}.
\label{S}
\eea
These quantities have nontrivial Lorentz-transformation properties.
For example,
the effective electron mass
$\tilde m$
depends on the observer inertial frame. 
Note that $\cC_{\mu 0}=0$,
which reflects the elimination 
of the time-derivative Lorentz-violating couplings
in the lagrangian \rf{lagc} for $\ch$.
Note also that $\cC_{\mu\nu}$ 
and hence $\tilde\et_{\mu\nu}$ are typically \it not \rm symmetric.
However,
$\cC_{\mu\nu}$ may be symmetric under certain circumstances,
such as in the special case of rotational invariance
in the chosen inertial frame,
where $\cC_{\mu\nu}$ becomes a diagonal matrix
with diagonal elements proportional to $(0,1,1,1)$.

To construct the relativistic quantum mechanics 
and the quantum field theory,
we follow the general procedure described
in Ref.\ \cite{kle}.
The modified Dirac equation for the free $\ch$ fermion
can be solved exactly using the plane-wave solution
\beq
\chi(x) = e^{-i \la_{\mu} x^{\mu}} w(\vec{\la}) .
\label{soln}
\eeq
Using the leading-order results \rf{redefc} and \rf{S},
the four-component spinor $w(\vec{\la})$ is found to satisfy 
\beq
(\tilde{\et}_{\mu\nu} \gamma^{\mu} \la^{\nu} - \tilde m ) w(\vec{\la}) 
= 0 .
\eeq
A nontrivial solution exists 
provided the determinant of the applied operator is zero.  
This leads to the dispersion relation 
\beq
\tilde{\la}^{2}  - \tilde{m}^{2} = 0 ,
\label{disp}
\eeq
where we define 
$\tilde{\la}_{\mu} \equiv \tilde{\et}_{\mu\nu} \la^{\nu}$.
This dispersion relation 
holds to leading order in $c_{\mu\nu}$.
It is quadratic in $\la^{0}(\vec{\la})$
and symmetric under $\la^{\mu} \rightarrow -\la^{\mu}$,
implying that
the roots are degenerate after reinterpretation.
There are therefore no energy splittings 
between the fermion and antifermion solutions
at this order,
and the dispersion relation
acquires only a momentum-dependent modification. 
We remark in passing that the exact dispersion relation for $\ch$
is identical to that for $\ps$ obtained from Eq.\ \rf{pslag}
and is given as Eq.\ (39) of Ref.\ \cite{kle},
where some of its properties are also discussed.
However,
the form \rf{disp} suffices for our purposes
and is convenient for calculation.

Define the positive root 
as $\la^{0}(\vec p) = p^0(\vec p) = E(\vec p)$,
where to leading order in $c_{\mu\nu}$ the energy $E$ is
\beq
E(\vec p) = 
\sqrt{\vec p^2 + \tilde m^2}
-\fr {p^j \cC_{jk} p^k}
{\sqrt{\vec p^2 + \tilde m^2}}
- \cC^0_{\pt{0}j} p^j.
\eeq
The field $\chi(x)$ is expanded as
\beq
\chi(x) = 
\int \fr {d^3 \vec{p}} {(2 \pi)^{3} N(\vec p)}
\sum_{\al = 1}^{2} \left[
b_{(\al)}(\vec p) e^{-i p \cdot x} u^{(\al)}(\vec p) 
+ d_{(\al)}^{\dagger}(\vec p) e^{i p \cdot x} v^{(\al)}(\vec p) 
\right] ,
\eeq
where the spinors are normalized to
\bea 
u^{(\al) \dagger} (\vec{p}) u^{(\al^{\prime})}(\vec{p}) =
\de^{\al \al^{\prime}} N(\vec p) 
& , & \quad
v^{(\al) \dagger} (\vec{p}) v^{(\al^{\prime})}(\vec{p}) =
\de^{\al \al^{\prime}} N(\vec p) ,
\nonumber \\
u^{(\al) \dagger} (\vec{p}) v^{(\al^{\prime})}(-\vec{p}) = 0
& , & \quad
v^{(\al) \dagger} (-\vec{p}) u^{(\al^{\prime})}(\vec{p}) = 0 .
\label{orthonorm}
\eea
We leave the normalization factor $N(\vec p)$ arbitrary 
to display explicitly its cancellation 
in the physical cross section.
The reader is cautioned
that boosting the chosen value of $N(\vec p)$
to another inertial frame is nontrivial 
because the spinor redefinition \rf{redefc}
must be taken into account.

Quantization is implemented by imposing the following nonvanishing 
anticommutation relations on the mode operators:
\bea
\{b_{(\al)} (\vec{p}), b^{\dagger}_{(\al^{\prime})}
(\vec{p}^{~\prime}) \} & = & (2 \pi)^3 
N(\vec p) 
\de_{\al \al^{\prime}}
\de^3 (\vec{p} - \vec{p}^{~\prime}) ,
\nonumber \\
\{d_{(\al)} (\vec{p}), d^{\dagger}_{(\al^{\prime})}
(\vec{p}^{~\prime}) \} & = & (2 \pi)^3 
N(\vec p)
\de_{\al \al^{\prime}}
\de^3 (\vec{p} - \vec{p}^{~\prime}) .
\eea
The resulting equal-time anticommutation relations 
for the $\ch$ field are conventional.
The corresponding single-particle states are 
therefore normalized as usual according to 
$\langle p^\prime,\al^\prime | p, \al \rangle 
= (2 \pi)^3 N(\vec p)
\de_{\al \al^\prime}
\de^3(\vec p^\prime - \vec p)$.
Matching this normalization to 
that obtained from the representation \rf{soln}
in relativistic quantum mechanics
shows that the number density 
for an incident plane wave is normalized to
$N(\vec p)$ particles per unit volume.

The conserved canonical energy-momentum tensor is 
an observer Lorentz 2-tensor,
which in terms of the redefined spinors $\ch$ in the chosen frame
takes the form 
\beq
\Theta^{\mn} = 
\fr i 2 
\tilde{\et}_{\al}^{\ \mu} 
\overline \chi 
\gamma^{\al} \lrprtnu \chi .
\eeq
The corresponding conserved four-momentum is diagonal
in the creation and annihilation operators
by virtue of the field redefinition:
\bea
P^{\mu} & = & \int d^{3} \vec x : \Theta^{0\mu} : 
\nonumber \\
& = & \int \fr {d^{3} \vec p} {(2 \pi)^{3} N(\vec p)} 
p^{\mu}\sum_{\al = 1}^{2} \left[ b_{(\al)}^{\dagger}(\vec p) 
b_{(\al)}(\vec p) + d_{(\al)}^{\dagger}(\vec p) d_{(\al)}(\vec p) 
\right] .
\eea
The conserved U(1) current is an observer Lorentz 4-vector
of the form
\beq
j^\mu = 
\tilde{\et}_\al^{\pt{\al}\mu}
\overline{\chi}
\ga^\al\chi ,
\eeq
with conserved charge 
\beq
Q = \int \fr {d^{3} \vec p} {(2 \pi)^{3} N(\vec p)} \sum_{\al = 
1}^{2} \left[ b_{(\al)}^{\dagger}(\vec p) b_{(\al)}(\vec p) 
- d_{(\al)}^{\dagger}(\vec p) d_{(\al)}(\vec p) \right] .
\eeq

To leading order in $c_{\mu\nu}$, 
the Feynman propagator \rf{genprop} for $\ch$ can be written as 
\beq
S_{F}(p) = \fr {1} {\tilde{p}_\mu\ga^\mu - \tilde{m}} .
\eeq
Equating the propagator to the vacuum expectation value 
of the usual time-ordered product of fields yields 
the useful identities
\beq
\sum_{\al = 1}^{2} u^{(\al)}(\vec p) \overline u^{(\al)}(\vec p)
= {N(\vec p) \over 2 \tilde E(\vec p)}({\slash{\tilde p}} + \tilde m) ,
\quad
\sum_{\al = 1}^{2} v^{(\al)}(\vec p) \overline v^{(\al)}(\vec p)
= {N(\vec p) \over 2 \tilde E(\vec p)}({\slash{\tilde p}} - \tilde m) ,
\label{identities}
\eeq
where 
$\tilde E(\vec p) \equiv \tilde p_{0}(\vec p) 
= \tilde \et_{0\nu} p^\nu$
and the shorthand notation ${\slash{\tilde p}} \equiv 
\tilde{\et}_{\mu\nu} \ga^{\mu} p^{\nu}$ is used.
These relations are generalizations of the usual ones,
and they can be verified by direct calculation using the
explicit expressions for the modified spinors.

\vglue 0.4cm
\noindent 
{\it 4.\ Cross section for $e^- e^+\to 2\ga$.}
As an illustrative example 
of a cross-section calculation in the QED extension,
we next apply the above general procedure 
to the pair annihilation of relativistic electrons and positrons
into two photons.
To simplify matters,
we disregard Lorentz-violating effects in the photon sector
associated with the coefficients
${(k_{AF})}_\mu$ and ${(k_F)}_{\ka\la\mu\nu}$.
The former is in any case expected theoretically to be zero
\cite{ck}
and is bounded experimentally 
to less than about $10^{-42}$ GeV
by astrophysical observations 
\cite{cfj},
while the latter is also known to be negligible 
on the relevant scale
\cite{ck,km}.

The appropriate element of the transition matrix $T$
for the process
$e^- e^+ \rightarrow 2\gamma$
can be found by evaluating 
the tree-level diagrams for electron-positron annihilation,
using the modified Feynman rules 
for the QED extension.
The result is
\bea
iT_{fi} &=& -ie^{2}(2\pi)^{4} \de^4 (k_1 + k_2 - p_1 - p_2) 
\overline{v}(p_{2}) \biggl[ {\slash{\tilde\ep_{2}}} \fr 1 
{{\slash{\tilde p_{1}}} - 
{\slash{\tilde k_{1}}} - \tilde m} {\slash{\tilde\ep_{1}}} + (1
\leftrightarrow 2)\biggr] u(p_{1}) 
\nonumber \\
&\equiv& i(2\pi)^{4} \de^{4} (k_1 + k_2 - p_1 - p_2) 
\cT_{fi}.
\eea
In this equation,
$p_1$, $p_2$ are the electron and positron momenta,
while $k_1$, $k_2$ are the photon momenta.
The spinors $u$, $v$ solve the modified Dirac equation
after the reinterpretation,
while $\ep_1$, $\ep_2$ are the two photon polarization vectors.
Note that crossing symmetry is satisfied,
as expected.

Following the discussions above,
it is most convenient 
to perform the whole calculation in a single inertial frame.
Since in practice the experimental procedure 
involves detecting back-to-back photons 
in reconstructing the cross section,
the appropriate frame is the center-of-momentum frame. 
This frame is also convenient for practical calculations.

The transition rate is determined by $\sum |\cT_{fi}|^2$,
where the sum is over 
final photon polarizations and initial fermion states.
In accordance with the relativistic approximation,
we treat $m$ as small 
and neglect subleading Lorentz-invariant factors
of order $m^2/|\vec p|^2$.
Taking advantage of the identities \rf{identities}
and performing algebraic steps similar to those for
the equivalent calculation in conventional QED
yields in the center-of-momentum frame 
\bea
\label{transprob}
\sum |\cT_{fi}|^{2} &=& 
{4e^4 N(\vec p)N(- \vec p)\over 
\tilde E(\vec p) \tilde E(- \vec p)} 
\biggl\{ 
\biggl(
\fr {1 + \cos^{2}{\theta}} 
{\sin^{2}{\theta}} 
\biggr) 
[1 - 2(c_{00} + c_{11} + c_{22} + c_{33})] 
\nonumber \\
&& 
\qquad \qquad
+ \fr {4} {\sin^{4}{\theta}}[\cos^{2}{\theta} 
\hat p c \hat p + \hat k c \hat k 
- \cos{\theta} (1 + \cos^{2}{\theta}) \hat k c \hat p]
\biggr\} ,
\label{t2}
\eea
which consists of the usual QED result
plus order-$c_{\mu\nu}$ corrections.
In this expression,
$\hat p$ and $\hat k$ are unit vectors 
along the incident electron and outgoing photon directions,
and $\cos\th =\hat p \cdot \hat k$. 
Terms such as $\hat k c \hat p$
denote double contraction 
of the matrix $c_{jk}$ with the indicated unit vectors.

\begin{figure} 
\centerline{\psfig{figure=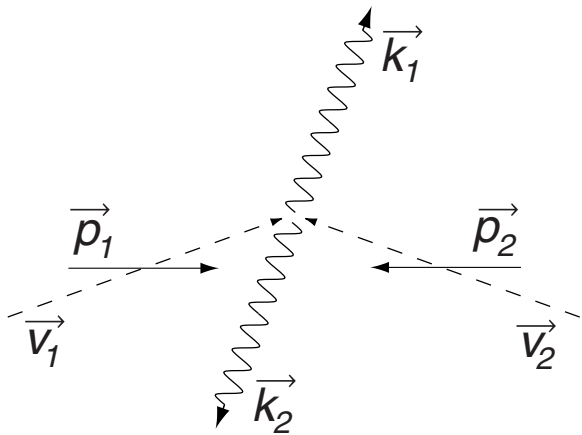,width=0.5\hsize}}
\smallskip
\caption{
Momenta and velocities for the process $e^- e^+\to 2\ga$.
The incoming fermion momenta 
$\vec p_1$, $\vec p_2$ are equal and opposite,
as are the outgoing photon momenta 
$\vec k_1$, $\vec k_2$ and velocities.
The incoming fermion group velocities $\vec v_1$, $\vec v_2$,
which determine the particle trajectories,
typically are \it not \rm parallel
by virtue of the modified velocity-momentum relationship
in Lorentz-violating quantum field theories. 
}
\label{fig1}
\end{figure}

The definition of the cross section also requires
the calculation of the factor $F$
in Eq.\ \rf{F}.
This must again be evaluated in the center-of-momentum frame.
The frame dependence of the spinor redefinition \rf{redef}
implies that it would,
for example,
be incorrect to proceed by calculating 
the flux and target density in the laboratory frame
and applying standard transformation laws
directly to boost the corresponding momenta back 
to the center-of-momentum frame.
This would generate spurious factors of $\ga^2$
that appear to dominate the cross section at high energies.

Keeping the normalizations arbitrary as before,
it suffices to determine the beam velocity difference.
Using Eqs.\ \rf{vg} and \rf{disp},
the group velocity is found to be 
\beq
v_g^j 
= {1 \over \tilde E}(\tilde p^j + \cC_\mu^{\pt{\mu}j} p^\mu) 
\label{velc}
\eeq
to lowest order in $\cC_{\mu\nu}$.
Note that this velocity typically does \it not \rm 
lie along the momentum $\vec p$.
Figure 1 illustrates the situation 
for the electron-positron pair annihilation.
In the center-of-momentum frame,
the momenta of the incoming beams are equal and opposite,
but the velocities (which determine the trajectories) 
typically are not.
For example,
in the present case we find 
$v_1^j +v_2^j = 2(c^{0j} + c^{j0})$,
which can be nonzero.

Using Eq.\ \rf{velc},
the velocity difference $|\vec v_1 - \vec v_2|$
can be deduced.
This in turn yields the factor $F$ in the center-of-momentum frame as
\beq
F = 2N(\vec p) N(- \vec p) (1 - \hat p \cC \hat p) .
\label{Fc}
\eeq
In this equation and what follows,
expressions such as $\hat p \cC \hat p$
are understood to denote double contraction
of the matrix $\cC_{jk}$
with the indicated unit vectors.
The reader is reminded that the beam velocity difference
(which is about 2 rather than 1)
is \it not \rm the relative beam velocity
obtained using a relativistic velocity-addition law.
Note also that Lorentz-symmetric contributions of order
$m^2/|\vec p|^2$ are neglected here
but would be important in any considerations of microcausality
\cite{kle}.

Combining the results for the transition rate and the factor $F$
given in Eqs.\ \rf{t2} and \rf{Fc},
using conventional phase-space factors for the final-state photons,
produces the cross section in the center-of-momentum frame:
\bea
{d \sigma \over d \Omega} &=& 
{\alpha^2 \over 4 \vec p^2}
\biggl\{
\biggl(
{1+\cos^2{\theta} \over \sin^2{\theta}}
\biggr) 
(1 + 3 \hat p \cC \hat p + 2 \cC_\mu^{\pt{\mu}\mu})
\nonumber \\ 
&&
\qquad\qquad
+ \fr {4} {\sin^{4}{\theta}}
[\cos^{2}{\theta} \hat p \cC \hat p 
+ \hat k \cC \hat k - \cos{\theta} (1 + \cos^{2}{\theta}) 
\hat k \cC \hat p] 
\biggr\} .
\label{xsect}
\eea
For definiteness,
we choose coordinates with $\vec p_1$ along the 3 axis.
Integrating this expression over the azimuthal angle $\ph$
then yields 
\bea
\fr {d\si}{d\cos\th} &=&
\int_{0}^{2 \pi} d \phi \fr {d\sigma}{d \Omega}
\nonumber\\
&=&  \fr {\pi \alpha^{2}} {2 \vec p^{2}}
\left[
\biggl(
\fr{1 + \cos^{2}{\theta}}
{\sin^{2}{\theta}}
\biggr)
(1+ c_{00} + c_{33})
-2 \cot^2{\theta}(c_{11} + c_{22} - 2 c_{33})
\right].
\nonumber\\
\label{xsectth}
\eea
Note that the normalization factors have cancelled.

For other processes,
fermions may be present in the final state. 
As usual,
the available phase space $d\Pi$ per final state fermion is
\beq
d\Pi =  \fr {d^3 \vec p} {(2 \pi)^3 N(\vec p)} .
\eeq
In observable cross sections or decay rates,
the normalization factors $N(\vec p)$ again 
cancel with those in the transition probability.
Note,
however,
that boosting $d\Pi$ to another inertial frame
is typically nontrivial
because the spinor redefinition \rf{redefc} is involved. 

\vglue 0.4cm
\noindent 
{\it 5.\ Experimental signals.}
The cross section \rf{xsect} is given in the 
center-of-momentum frame,
and the components of the coefficients $c_{\mu\nu}$ 
for Lorentz violation are also defined in this frame.
For symmetric colliders,
the center-of-momentum frame coincides
with the laboratory frame.
However,
the laboratory frame rotates with the Earth,
so the spatial components of $c_{\mu\nu}$ 
oscillate periodically as functions of the sidereal time $t$.
This induces corresponding variations
in the observed cross section,
with periodicities controlled by the Earth's sidereal
rotation frequency $\Om \simeq 2\pi$/(23 h 56m).

To display explicitly the time dependence
of the observable cross section,
we introduce suitable bases of vectors
for the nonrotating frame and the laboratory frame,
following the notation and conventions of Ref.\ \cite{kla}.
In this choice of coordinates,
the right-handed basis
$(\hat X, \hat Y, \hat Z )$
for the nonrotating frame
is compatible with celestial equatorial coordinates
\cite{celestial}
with $\hat Z$ aligned along the rotation axis of the Earth.
The other basis vectors lie in the plane of the equator,
with $\hat X$ having declination and right ascension $0^\circ$,
and $\hat Y$ having declination $0^\circ$ and right ascension $90^\circ$.
Although this coordinate system is in fact only approximately fixed 
because the Earth precesses slowly over time, 
any induced effects are suppressed by several orders of magnitude 
and are neglected here.  

We select the basis $(\hat x, \hat y, \hat z )$
in the laboratory frame
such that $\hat z \equiv \hat p_1$
is aligned along the direction of the electron-beam momentum.
The vector $\hat x$ is fixed 
by requiring it to be perpendicular 
to $\hat z$ and to lie in the $\hat z$-$\hat Z$ plane.
The remaining vector $\hat y$ is chosen to complete
a right-handed triad: 
$\hat y = \hat z \times \hat x$.
The angle between $\hat z$ and $\hat Z$
is denoted by $\ch$, so $\hat z \cdot \hat Z = \cos \ch$. 

The transformation between the two sets of bases
can be regarded as nonrelativistic 
to an excellent approximation.
It is given by
\beq 
\left(
\begin{array}{c}
\hat x \\
\hat y \\
\hat z
\end{array}
\right)
= 
\left(
\begin{array}{ccc}
\cos{\chi} \cos{\Omega t} & \cos{\chi} \sin{\Omega t} &
- \sin{\ch} \\
- \sin{\Omega t} & \cos{\Omega t} & 0 \\
\sin{\chi} \cos{\Omega t} & \sin{\chi} \sin{\Omega t} & 
\cos{\ch} 
\end{array}
\right) 
\left(
\begin{array}{c}
\hat X \\
\hat Y \\
\hat Z
\end{array}
\right).
\label{transform}
\eeq
In what follows,
we denote indices on the coefficients for Lorentz violation 
in the laboratory frame
by $0, 1, 2, 3$
and indices in the nonrotating frame by 
$0, X, Y, Z$.

To exhibit the sidereal-time dependence 
of the cross section,
it suffices to apply the transformation \rf{transform}
to the coefficients $c_{\mu\nu}$ for Lorentz violation.
For example,
the combinations of coefficients appearing 
in the cross section \rf{xsectth}
become
\bea
\hbox{\hskip-20pt}
c_{11} + c_{22} - 2 c_{33} &=& 
(c_{XX} + c_{YY} - 2 c_{ZZ})
(\frac 3 2 \cos^2{\chi} - \half)
\nonumber \\ 
&&
- \frac 3 2 (c_{XZ} + c_{ZX}) \sin{2 \chi} \cos{\Omega t}
- \frac 3 2 (c_{YZ} + c_{ZY}) \sin{2 \chi} \sin{\Omega t}
\nonumber \\
&&
- \frac 3 2 (c_{XX} - c_{YY}) \sin^2{\chi} \cos{2 \Omega t}
- \frac 3 2 (c_{XY} + c_{YX}) \sin^2{\chi} \sin{2 \Omega t} ,
\nonumber\\
c_{00} + c_{33} &=& 
c_{00} + c_{ZZ} 
+ \half (c_{XX} + c_{YY} - 2 c_{ZZ}) \sin^2{\chi} 
\nonumber \\
&&
+ \half (c_{YZ} + c_{ZY})\sin{2\ch} \sin{\Omega t} 
+ \half (c_{XZ} + c_{ZX})\sin{2\ch} \cos{\Omega t}
\nonumber \\
&& 
+ \half (c_{XY} + c_{YX})\sin^2{\ch} \sin{2\Omega t}
+ \half (c_{XX} - c_{YY})\sin^2{\ch} \cos{2\Omega t} .
\nonumber\\
\label{combs}
\eea
These endow the observable cross section 
with a dependence on sidereal time $t$
that has three components:
a constant,
an oscillation with period $T = 2\pi/\Om$,
and an oscillation with period $T/2$.

Data in experiments are usually taken 
at different sidereal times and over several months.
A typical analysis searching for deviations from QED
in the cross section for $e^-e^+$ annihilation
would therefore effectively average away
the sidereal-time dependence.
With coefficients expressed
in the nonrotating coordinate system,
the typical analysis is thus sensitive
only to the average
\bea
\fr {d\si}{d\cos\th}\biggr\vert _{\rm av} &\equiv&
\fr 1 T \int_{0}^{T} dt \fr {d\si}{d\cos\th}
\nonumber\\
&=& 
\fr {\pi\alpha^{2}} {2 \vec p^{2}}
\biggl\{ 
\biggl(
\fr {1 + \cos^{2}{\th}} {\sin^{2}{\th}}
\biggr)
\Big[
1 + c_{00} + c_{ZZ}
+ \half (c_{XX} + c_{YY} - 2 c_{ZZ})\sin^2{\chi} 
\Big]
\nonumber\\ 
&& 
\qquad
- 2 \cot^2{\theta} (c_{XX} + c_{YY} - 2 c_{ZZ})
(\frac 3 2 \cos^2{\chi} - \frac 1 2)
\biggr\} .
\eea
This shows the time-averaged effect of the Lorentz-violating terms 
is the sum of a scaling of the usual cross section 
with a correction term proportional to $\cot^2{\theta}$.

The observation of sidereal variations in cross sections 
would provide a unique signal of Lorentz violation.
In high-energy physics,
analogous effects for neutral-meson oscillations
have been used by the KTeV Collaboration
to obtain a new constraint on coefficients for CPT violation
in the standard-model extension
\cite{k99}.
Sidereal variations also form the basis for a variety
of high-precision low-energy tests of
the standard-model extension
\cite{rm,dp,vh,lh,db,bh}.
However,
to date constraints have been placed only on
a few of the coefficients $c_{\mu\nu}$.

Large data sets of several hundred inverse picobarns
for relativistic electron-positron pair annihilation
have recently been collected by experiments at LEP
\cite{lepexpts}.
These could in principle be used to bound the combinations 
of the coefficients $c_{\mu\nu}$ 
in the nonrotating frame
that appear in Eqs.\ \rf{combs}.
Note that the Lorentz-violating modifications 
to the cross section \rf{xsectth}
show no enhancement with energy, 
in accordance with the discussion 
at the beginning of section 3.
This contrasts with corrections to the conventional QED
cross section arising from other types of new physics,
which typically grow with energy
\cite{enn}.
Although the necessary analysis and the systematics 
are qualitatively different in nature from those performed to date,
it seems unlikely that 
sidereal-time binning of the $e^-e^+\to2\ga$ data
would yield tight bounds.
However,
it could limit certain components of $c_{\mu\nu}$
otherwise presently unconstrained by experiment.

\vglue 0.4cm
\noindent 
{\it 6.\ Summary.}
In this work,
a procedure is presented 
for calculating observable cross sections and decay rates 
in the Lorentz-violating standard-model extension.
In determining the transition matrix,
conventional perturbative techniques apply
but with modified Feynman rules.
To avoid complexities associated with 
inertial-frame changes in the presence of Lorentz violation,
it is most convenient to perform any analysis
in a single frame.
Methods for obtaining the necessary kinematical factors 
are also presented.

As an example,
the process $e^-e^+ \to 2 \ga$
for relativistic electrons and positrons
is explicitly considered.
In this case,
the relevant coefficients for Lorentz violation
are associated with derivative couplings in the 
fermion sector of the Lorentz-violating QED extension,
and they scale with momentum like the usual fermion kinetic term.
The contributions from other coefficients either are negligible 
or average to zero for unpolarized scattering.
The resulting cross section is explicitly given 
in Eqs.\ \rf{xsect} and \rf{xsectth}. 
The Earth's rotation induces a dependence 
of this cross section on sidereal time,
specified in Eq.\ \rf{combs}.
High-statistics experiments searching for this effect
could place bounds on some coefficients for Lorentz violation
otherwise unconstrained by experiment. 

It would be of interest to obtain expressions 
for other standard QED scattering amplitudes
in the context of the Lorentz-violating QED extension.
It is possible that the experimental sensitivity
is enhanced by the boost factor $\ga$ in appropriate circumstances.
Certainly,
sensitivity to different parameters can be expected
among different processes. 
Also,
the above arguments for neglecting 
certain coefficients for Lorentz violation
fail for nonrelativistic fermions,
so different scenarios such as fixed-target experiments
could be worth investigation.

\vglue 0.4cm
We thank Salvatore Mele for discussion.
This work was supported in part
by the United States Department of Energy 
under grant DE-FG02-91ER40661.

\vglue 0.4cm


\begin{thebibliography}{xx}

\baselineskip=19pt

\bibitem{cpt98}
For overviews of various theoretical ideas
and experimental tests of Lorentz symmetry,
see, for example,
V.A.\ Kosteleck\'y, ed.,
\it CPT and Lorentz Symmetry, \rm
World Scientific, Singapore, 1999.

\bibitem{ck} 
D.\ Colladay and V.A.\ Kosteleck\'y,
Phys.\ Rev.\ D {\bf 55}, 6760 (1997);
Phys.\ Rev.\ D {\bf 58}, 116002 (1998).

\bibitem{kle} 
V.A.\ Kosteleck\'y and R.\ Lehnert,
Phys.\ Rev.\ D {\bf 63}, 065008 (2001).

\bibitem{bkr} 
R.\ Bluhm
{\it et al.},
Phys.\ Rev.\ Lett.\ {\bf 79}, 1432 (1997);
Phys.\ Rev.\ D {\bf 57}, 3932 (1998).

\bibitem{gg}
G.\ Gabrielse 
{\it et al.},
in Ref.\ \cite{cpt98};
Phys.\ Rev.\ Lett.\ {\bf 82}, 3198 (1999).

\bibitem{hd}
H.\ Dehmelt 
{\it et al.},
Phys.\ Rev.\ Lett.\ {\bf 83}, 4694 (1999).

\bibitem{rm}
R.\ Mittleman, I.\ Ioannou, and H.\ Dehmelt,
in Ref.\ \cite{cpt98};
R.\ Mittleman 
{\it et al.},
Phys.\ Rev.\ Lett.\ {\bf 83}, 2116 (1999).

\bibitem{bkr2} 
R.\ Bluhm
{\it et al.},
Phys.\ Rev.\ Lett.\ {\bf 82}, 2254 (1999).

\bibitem{dp}
D.F.\ Phillips
{\it et al.},
Phys.\ Rev.\ D {\bf 63}, 111101 (2001).

\bibitem{bkl} 
R.\ Bluhm
{\it et al.},
Phys.\ Rev.\ Lett.\ {\bf 84}, 1098 (2000).

\bibitem{vh} 
V.W.\ Hughes
{\it et al.},
presented at the Hydrogen II Conference,
Tuscany, Italy,
June, 2000.

\bibitem{ccexpt}
V.W.\ Hughes, H.G.\ Robinson, and V.\ Beltran-Lopez,
Phys.\ Rev.\ Lett.\ {\bf 4}, 342 (1960);
R.W.P.\ Drever,
Philos.\ Mag.\ {\bf 6}, 683 (1961);
J.D.\ Prestage 
{\it et al.},
Phys.\ Rev.\ Lett.\ {\bf 54}, 2387 (1985);
S.K.\ Lamoreaux {\it et al.},
Phys.\ Rev.\ Lett.\ {\bf 57}, 3125 (1986);
Phys.\ Rev.\ A {\bf 39}, 1082 (1989);
T.E.\ Chupp
{\it et al.},
Phys.\ Rev.\ Lett.\ {\bf 63}, 1541 (1989);
C.J.\ Berglund
{\it et al.},
Phys.\ Rev.\ Lett.\ {\bf 75}, 1879 (1995).

\bibitem{kla}
V.A.\ Kosteleck\'y and C.D.\ Lane,
Phys.\ Rev.\ D {\bf 60}, 116010 (1999);
J.\ Math.\ Phys.\ {\bf 40}, 6245 (1999).

\bibitem{lh}
L.R.\ Hunter
{\it et al.},
in Ref.\ \cite{cpt98}.

\bibitem{db}
D.\ Bear
{\it et al.},
Phys.\ Rev.\ Lett.\ {\bf 85}, 5038 (2000).

\bibitem{bk} 
R.\ Bluhm and V.A.\ Kosteleck\'y,
Phys.\ Rev.\ Lett.\ {\bf 84}, 1381 (2000).

\bibitem{bh} 
B.\ Heckel
{\it et al.},
in B.N.\ Kursunoglu, S.L.\ Mintz, and A.\ Perlmutter, eds.,
\it Elementary Particles and Gravitation, \rm
Plenum, New York, 1999.

\bibitem{cfj}
S.M. Carroll, G.B. Field, and R. Jackiw, 
Phys. Rev. D {\bf 41}, 1231 (1990).

\bibitem{jk}
R.\ Jackiw and V.A.\ Kosteleck\'y,
Phys.\ Rev.\ Lett.\ {\bf 82}, 3572 (1999).

\bibitem{pvc}
M.\ P\'erez-Victoria, 
Phys.\ Rev.\ Lett.\ {\bf 83}, 2518 (1999);
J.M.\ Chung, 
Phys.\ Lett.\ B {\bf 461}, 138 (1999).

\bibitem{kpcvk}
V.A.\ Kosteleck\'y and R.\ Potting,
Phys.\ Rev.\ D {\bf 51}, 3923 (1995);
D.\ Colladay and V.A.\ Kosteleck\'y,
Phys.\ Lett.\ B {\bf 344}, 259 (1995);
Phys.\ Rev.\ D {\bf 52}, 6224 (1995);
V.A.\ Kosteleck\'y and R.\ Van Kooten,
Phys.\ Rev.\ D {\bf 54}, 5585 (1996);
V.A.\ Kosteleck\'y,
Phys.\ Rev.\ Lett.\ {\bf 80}, 1818 (1998);
Phys.\ Rev.\ D {\bf 61}, 016002 (2000);
hep-ph/0104120;
N.\ Isgur
{\it et al.},
in preparation.

\bibitem{k99}
KTeV Collaboration,
Y.B.\ Hsiung 
{\it et al.},
Nucl.\ Phys.\ Proc.\ Suppl.\ {\bf 86}, 312 (2000).
 
\bibitem{bexpt}
OPAL Collaboration, 
R.\ Ackerstaff 
{\it et al.},
Z.\ Phys.\ C {\bf 76}, 401 (1997);
DELPHI Collaboration,
M.\ Feindt 
{\it et al.},
preprint DELPHI 97-98 CONF 80 (July 1997);
BELLE Collaboration,
K.\ Abe
{\it et al.},
Phys.\ Rev.\ Lett.\ {\bf 86}, 3228 (2001).

\bibitem{bckp}
O.\ Bertolami
{\it et al.},
Phys.\ Lett.\ B {\bf 395}, 178 (1997).

\bibitem{kps}
V.A.\ Kosteleck\'y and S.\ Samuel,
Phys.\ Rev.\ D {\bf 39}, 683 (1989);
{\it ibid.} 
{\bf 40}, 1886 (1989);
Phys.\ Rev.\ Lett.\ {\bf 63}, 224 (1989);
{\it ibid.} 
{\bf 66}, 1811 (1991);
V.A.\ Kosteleck\'y and R.\ Potting,
Nucl.\ Phys.\ B {\bf 359}, 545 (1991);
Phys.\ Lett.\ B {\bf 381}, 89 (1996);
Phys.\ Rev.\ D {\bf 63}, 046007 (2001); 
V.A.\ Kosteleck\'y, M.\ Perry, and R.\ Potting,
Phys.\ Rev.\ Lett.\ {\bf 84}, 4541 (2000). 

\bibitem{km}
V.A.\ Kosteleck\'y and M.\ Mewes,
in preparation.

\bibitem{celestial}
R.M.\ Green,
\it Spherical Astronomy, \rm
Cambridge University Press, Cambridge, 1985.

\bibitem{lepexpts}
ALEPH Collaboration,
R.\ Barate
{\it et al.},
Phys.\ Lett.\ B {\bf 429}, 201 (1998);
DELPHI Collaboration,
P.\ Abreu 
{\it et al.},
Phys.\ Lett.\ B {\bf 433}, 429 (1998);
Phys.\ Lett.\ B {\bf 491}, 67 (2000);
L3 Collaboration,
M.\ Acciarri 
{\it et al.},
Phys.\ Lett.\ B {\bf 413}, 159 (1997);
Phys.\ Lett.\ B {\bf 475}, 198 (2000);
OPAL Collaboration,
K.\ Ackerstaff
{\it et al.},
Eur.\ Phys.\ J.\ C {\bf 1}, 21 (1998);
Phys.\ Lett.\ B {\bf 438}, 379 (1998);
Phys.\ Lett.\ B {\bf 465}, 303 (1999).

\bibitem{enn}
O.J.P\ \'Eboli, A.A.\ Natale, and S.F.\ Novaes,
Phys.\ Lett.\ B {\bf 271}, 274 (1991).

\end{thebibliography}
\end{document}